\documentclass[sigconf]{acmart}
\usepackage{CJKutf8}
\newcommand{\cgray}[1]{\textcolor{gray}{#1}}
\newcommand{\cblue}[1]{\textcolor{blue}{#1}}
\newcommand{\cred}[1]{\textcolor{red}{#1}}
\usepackage{multirow}
\usepackage{makecell}
\usepackage{graphicx}

\AtBeginDocument{%
  }

\copyrightyear{2025}
\acmYear{2025}
\setcopyright{acmlicensed}\acmConference[KDD '25]{Proceedings of the 31st ACM SIGKDD Conference on Knowledge Discovery and Data Mining V.2}{August 3--7, 2025}{Toronto, ON, Canada}
\acmBooktitle{Proceedings of the 31st ACM SIGKDD Conference on Knowledge Discovery and Data Mining V.2 (KDD '25), August 3--7, 2025, Toronto, ON, Canada}
\acmDOI{10.1145/3711896.3737265}
\acmISBN{979-8-4007-1454-2/2025/08}

\begin{document}
\begin{CJK}{UTF8}{song}

\title{Synergizing Implicit and Explicit User Interests: A Multi-Embedding Retrieval Framework at Pinterest}

\author{Zhibo Fan}
\email{zb1439@outlook.com}
\affiliation{%
  \institution{Pinterest}
  \city{San Francisco}
  \state{CA}
  \country{USA}
}
\authornote{Work done at Pinterest.}
\author{Hongtao Lin}
\email{hongtaolin@pinterest.com}
\affiliation{%
  \institution{Pinterest}
  \city{San Francisco}
  \state{CA}
  \country{USA}
}
\author{Haoyu Chen}
\email{hchen@pinterest.com}
\affiliation{%
  \institution{Pinterest}
  \city{San Francisco}
  \state{CA}
  \country{USA}
}
\authornotemark[1]
\author{Bowen Deng}
\email{bdeng@pinterest.com}
\affiliation{%
  \institution{Pinterest}
  \city{San Francisco}
  \state{CA}
  \country{USA}
}
\author{Hedi Xia}
\email{hxia@pinterest.com}
\affiliation{%
  \institution{Pinterest}
  \city{San Francisco}
  \state{CA}
  \country{USA}
}
\authornote{Corresponding author.}
\author{Yuke Yan}
\email{yukeyan@pinterest.com}
\affiliation{%
  \institution{Pinterest}
  \city{San Francisco}
  \state{CA}
  \country{USA}
}
\author{James Li}
\email{jamesyili@gmail.com}
\affiliation{%
  \institution{Pinterest}
  \city{San Francisco}
  \state{CA}
  \country{USA}
}

% \author{Zhibo Fan$^{\dagger}$, Hongtao Lin, Haoyu Chen$^{\dagger}$, Bowen Deng, Hedi Xia, Yuke Yan, James Li}
% \email{{zhibofan, hongtaolin, hchen, bdeng, hxia, yukeyan, jamesli}@pinterest.com}
% \affiliation{%
%   \institution{Pinterest}
%   \state{California}

\renewcommand{\shortauthors}{Zhibo Fan et al.}

\begin{abstract}
  Industrial recommendation systems are typically composed of multiple stages, including retrieval, ranking, and blending. The retrieval stage plays a critical role in generating a high-recall set of candidate items that covers a wide range of diverse user interests. Effectively covering the diverse and long-tail user interests within this stage poses a significant challenge: traditional two-tower models struggle in this regard due to limited user-item feature interaction and often bias towards top use cases. To address these issues, we propose a novel multi-embedding retrieval framework designed to enhance user interest representation by generating multiple user embeddings conditioned on both implicit and explicit user interests. Implicit interests are captured from user history through a Differentiable Clustering Module (DCM), whereas explicit interests, such as topics that the user has followed, are modeled via Conditional Retrieval (CR). These methodologies represent a form of conditioned user representation learning that involves condition representation construction and associating the target item with the relevant conditions. Synergizing implicit and explicit user interests serves as a complementary approach to achieve more effective and comprehensive candidate retrieval as they benefit on different user segments and extract conditions from different but supplementary sources. Extensive experiments and A/B testing reveal significant improvements in user engagements and feed diversity metrics. Our proposed framework has been successfully deployed on Pinterest home feed.

% \begingroup
% \renewcommand\thefootnote{$\dagger$}
% \footnotetext{Work done at Pinterest.}
% \endgroup
\end{abstract}




% \received{03 February 2025}
% \received[accepted]{16 May 2025}

\maketitle

%%%%%%%%%%%%%%%%%%%%%%%%%%%%%%%
%%%%%%%%%%%%%%%%%%%%%%%%%%%%%%%
\section{Introduction}
\begin{figure}[tbp]
    \centering
    \includegraphics[width=0.95\linewidth]{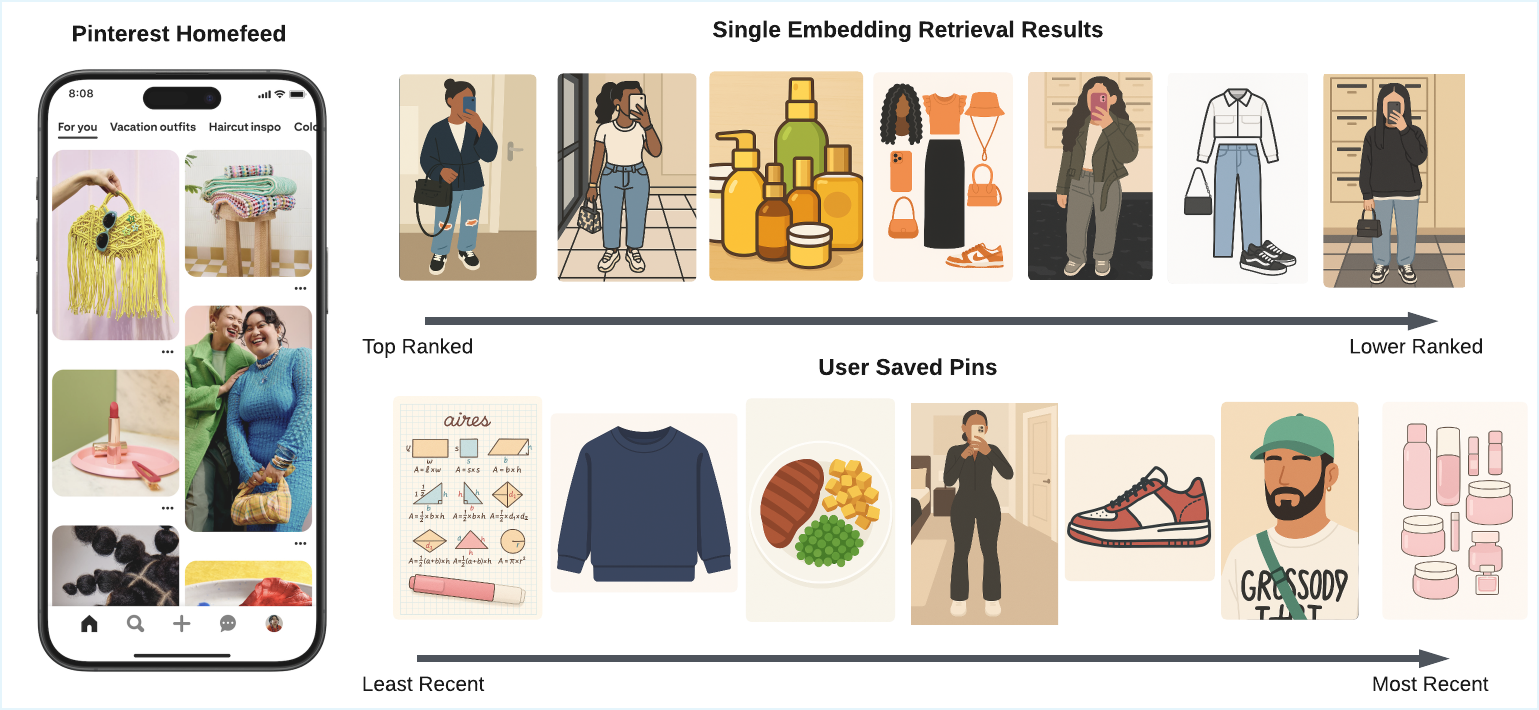}
    \caption{Left: Pinterest home feed screenshot. Right: A failure case of a vanilla two-tower model. The top retrieved candidates from the two-tower model (above) fail to cover user interests, such as food and education, from the user's previously saved Pins (below). Images from user history and recommendation results are iconized to obscure identifiable user information.}
    \label{fig:homefeed}
\end{figure}

%ACM's consolidated article template, introduced in 2017, provides a
%consistent \LaTeX\ style for use across ACM publications, and
%incorporates accessibility and metadata-extraction functionality
%necessary for future Digital Library endeavors. Numerous ACM and
%SIG-specific \LaTeX\ templates have been examined, and their unique
%features incorporated into this single new template.
%
%If you are new to publishing with ACM, this document is a valuable
%guide to the process of preparing your work for publication. If you
%have published with ACM before, this document provides insight and
%instruction into more recent changes to the article template.
%
%The ``\verb|acmart|'' document class can be used to prepare articles
%for any ACM publication --- conference or journal, and for any stage
%of publication, from review to final ``camera-ready'' copy, to the
%author's own version, with {\itshape very} few changes to the source.
%
%\section{Template Overview}
%As noted in the introduction, the ``\verb|acmart|'' document class can
%be used to prepare many different kinds of documentation --- a
%double-anonymous initial submission of a full-length technical paper, a
%two-page SIGGRAPH Emerging Technologies abstract, a ``camera-ready''
%journal article, a SIGCHI Extended Abstract, and more --- all by
%selecting the appropriate {\itshape template style} and {\itshape
%  template parameters}.
%
%This document will explain the major features of the document
%class. For further information, the {\itshape \LaTeX\ User's Guide} is
%available from
%\url{https://www.acm.org/publications/proceedings-template}.
As one of the largest visual discovery platforms, Pinterest hosts a billion-scale visual content gallery and inspires over 500 million users worldwide. Upon visiting Pinterest, users are immediately presented with a diversified and inspiring home feed (Figure \ref{fig:homefeed}, left) designed to help them discover new ideas. To enhance user engagement, it is crucial to surface recommendations that capture multiple user interests at first glance.

Pinterest home feed follows a modern multi-stage recommendation system architecture, consisting of retrieval, ranking, and blending. The retrieval stage focuses on generating candidate items with high recall and low serving latency. It sets the upper bound for the quality of recommendations, as subsequent stages primarily refine and re-rank the retrieved results. Different from Pin search~\cite{jing2015visual} and related Pins~\cite{liu2017related}, home feed retrieval often lacks explicit user browsing intents or contextual cues. Therefore, the key to effective home feed retrieval lies in holistically understanding user interests and ensuring sufficient coverage of relevant content.

Among various retrieval algorithms and strategies, embedding-based retrieval~\cite{shen2014learning-dssm,covington2016deep-jay} has been proved to be one of the most effective and widely adopted approaches in industry, where both users and candidate items are encapsulated as latent embedding vectors, with dot-product (or cosine similarity) as the affinity measure. Such a formulation facilitates efficient retrieval via approximate nearest neighbor search (ANNS)~\cite{malkov2018efficient-hnsw,li2019approximate}. These models are commonly referred to as "Two-Tower Models" \cite{shen2014learning-dssm,covington2016deep-jay,yi2019sampling} due to their architecture, where user and item information are encoded separately in a user tower and an item tower, without interaction before the similarity computation. While effective and widely used, Two-Tower Models can struggle to capture long-tail user interests due to the lack of early-stage feature interaction between user and item features. Empirically, we observed that dominant user interests tend to overwhelm the user embedding, leading to fewer retrieved candidates from the torso and tail of a user's interest distribution. Figure \ref{fig:homefeed} (right) illustrates an example where a single-embedding retrieval model overlooks user interests. This contradicts the primary objective of the retrieval stage--to maximize user interest coverage--and can negatively impact user engagement and retention.

\begin{figure}
    \centering
    \includegraphics[width=\linewidth]{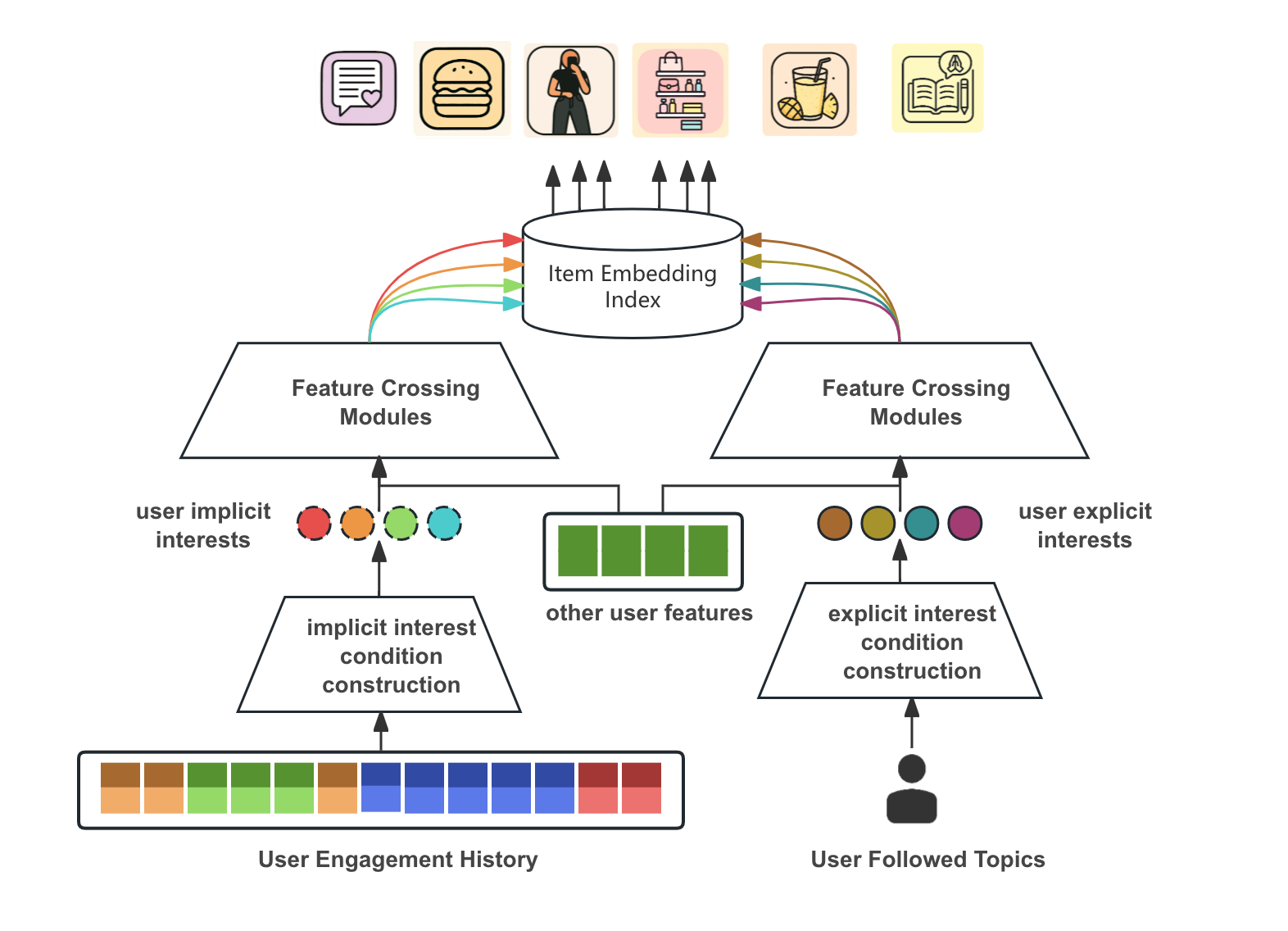}
    \caption{The multi-embedding retrieval framework conditioned on both implicit and explicit user interests.% Implicit interests are extracted from user sequence and explicit interests are selected by users. Both sources are used as conditions to generate multiple user-interest embeddings for candidate retrieval in parallel.
    }
    \label{fig:framework}
\end{figure}

To address this, we propose a multi-embedding retrieval framework (see Figure \ref{fig:framework}) to guide user embedding generation with both implicit and explicit user interests as conditions. It consists of  (A) an implicit interest conditioned model utilizing a Differentiable Clustering Module (DCM) to extract implicit interests from user historical engagements, and (B) a conditional retrieval (CR) \cite{lin2024bootstrapping} model to retrieve candidate items related to explicit user interests, such as followed topics and long-term interests that are otherwise "forgotten" due to user history input limit. 
Both models generate multiple embeddings per user to represent diverse interests, but each of the models captures a different aspect of user interests to complement the other: implicit interest modeling quickly adapts to users' realtime engagements and benefits active users the most, 
whereas explicit interest conditioned retrieval not only replenishes overlooked or long-term interests for active users, but also acts as an important candidate source for new and low-signal users. Synergizing both models improves user interest coverage and drives higher online engagement across all user segments. 

Conceptually, both models can be framed as conditional representation learning, involving two key aspects: 
\begin{enumerate}
    \item \textbf{Condition construction}: user interests are extracted and encoded into embeddings, and
    \item \textbf{Condition association}: engaged candidates are linked to the corresponding interests to form training labels.
\end{enumerate}
Specifically, DCM applies differentiable clustering over the user sequence to form latent user interest embeddings as conditions and associate them with positive labels inside the neural network. On the other hand, CR keeps record of source interests for engaged items at logging time and learns a topic embedding table to encode the conditions.

In summary, our main contributions of this work are as follows.
\begin{itemize}
\item We build a multi-embedding retrieval framework at Pinterest, which improves user interest coverage and provides highly engaging and diversified contents by synergizing both implicit and explicit user interests.
\item We present our solution for implicit interest modeling, namely DCM, together with our extensive survey on a series of SoTA modeling strategies, and show the superior performance of DCM among others. 
\item We deploy the framework at Pinterest and verify its effectiveness via extensive online experiments. The deployed system improves major user engagement metrics by a large margin.
\end{itemize}

%%%%%%%%%%%%%%%%%%%%%%%%%%%%%%%
%%%%%%%%%%%%%%%%%%%%%%%%%%%%%%%
\section{Related Works}\label{sec:related}

% \subsection{Retrieval and Two-Tower Models}
% Deep learning has revolutionized recommendation systems by enabling more effective modeling of user-item interaction\cite{covington2016deep-jay,shen2014learning-dssm}. Many methods proposed in the domain focus on feature interaction\cite{wang2021dcn,wang2021masknet,zhang2022dhen}, sequence modeling\cite{zhou2018deep-din,kang2018self-sasrec,pi2020search-sim,li2019multi-mind,xia2023transact,chang2023twin}, and representation learning\cite{shen2014learning-dssm,ying2018graph-pinsage,pancha2022pinnerformer,agarwal2024omnisearchsage,borisyuk2024lignn}. As a widely adopted retrieval modeling approach, Two-Tower Models\cite{shen2014learning-dssm,yi2019sampling} produce an embedding for user and item and use dot-product to measure the similarity or action probability between them.

% \subsection{Multi-Interest Retrieval}
In this section, we focus on discussing prior work on \textbf{embedding-based multi-interest retrieval} that is most relevant to our approach. We roughly categorize existing methods into three streams.

\textbf{Self-Attention.} This stream uses multi-head self-attention~\cite{vaswani2017attention,kang2018self-sasrec,sun2024coactiongraphrec} to derive multiple user interests from the user engagement sequence. Rather than aggregating the multi-head outputs to form a single user embedding, these methods\cite{xiao2020dmin,cen2020controllable-comirec} treat the output of each attention head as a distinct embedding to represent one of the user's latent interests.

\textbf{Interest Token.} Another line of research~\cite{wang2023incremental-limarec,tan2021sparse-sine,ma2019learning-macrid,wang2019modeling-mcprn,xu2022mixture-mvke,liu2024kuaiformer} involves learnable model parameters to represent a range of interests explicitly. SINE~\cite{tan2021sparse-sine} initializes a set of learnable concept prototype embeddings and activate top-K concepts per user request to attend to the user sequence. MVKE~\cite{xu2022mixture-mvke} introduces a set of "virtual kernels" to bridge the user tower and item tower. Omitting the multi-task setup, these "virtual kernels" can be interpreted as learnable query interest tokens for attention over user sequence. From this perspective, attention head weights in self-attentive methods~\cite{xiao2020dmin,cen2020controllable-comirec} can also be viewed as a special form of interest tokens. Additionally, rather than learning the concepts and their representations jointly, Conditional Retrieval (CR)~\cite{lin2024bootstrapping} proposes to learn the embeddings for predefined conditions to provide better controllability when "bootstrapping for new use cases". In this paper, we also adopted CR for explicit interest modeling to complement user interest coverage. 

\textbf{Dynamic Routing.} Unlike other modeling streams, dynamic routing based approaches~\cite{li2019multi-mind,chai2022user-umi,wang2020disentangled} do not introduce specific model parameters to represent interests. Instead, they derive interest representations from diverse patterns within the user sequence via dynamic routing~\cite{sabour2017dynamic-capsule}, which enables personalized interest granularity based on the varying diversity of user history. MIND~\cite{li2019multi-mind} applies Capsule Networks~\cite{sabour2017dynamic-capsule} to extract the interest representations from the user sequence and UMI~\cite{chai2022user-umi} enhances the routing by incorporating extra attention. Our deployed implicit interest modeling solution follows this line of research.

In addition to these modeling techniques, there are also works establishing multi-interest retrieval frameworks. ComiRec~\cite{cen2020controllable-comirec} extends self-attention and MIND~\cite{li2019multi-mind} with a controllable diversification layer to build an implicit user interest modeling framework. Trinity~\cite{yan2024trinity} devises a complex system that extracts explicit user interests from different sources with machine learning and heuristics. In this paper, we present a deployed framework utilizing both implicit and explicit user interest models to cover different sources of user interests and boost  the performance across all user segments.

\section{Method}
In this section, we elaborate on our proposed multi-embedding retrieval framework, starting from a problem statement as an overview of the framework. Next, we present the best performing solution, DCM, along with other alternatives for implicit interest modeling with an emphasis on how condition construction and association are implemented. After that, we describe the explicit interest modeling with CR~\cite{lin2024bootstrapping}. Finally, we summarize the connection and the supplementary effect of the framework components and present deployment details.

\subsection{Problem Statement}
The general objective of recommendation models is to predict the ranking score for an item $i$ given the user information $u$. Specifically, for retrieval stage models, to facilitate efficient retrieval, the score is usually formulated as $f(i|u)\propto \exp(\phi(u)^\top\psi(i))$, where $\phi$ and $\psi$ are neural functions, aka "towers", that map input features of $i$ and $u$ to lower dimensional normalized vectors. Without cross-tower feature interactions, such a formulation tends to overamplify the head user interests in $\phi(u)$ and result in worse coverage for torso and tail user interests.

Our proposed multi-embedding framework introduces $K_{im}$ implicit and $K_{ex}$ explicit user interest conditions to guide the user embedding generation as shown in Figure \ref{fig:framework}, where $K_{im}$ and $K_{ex}$ are pre-defined parameters. For a user $u$ with interest condition $c$, we have more information to tell if they will engage with an item $i$ and thus could revamp the ranking score as $f(i|u,c)\propto \exp(\phi(u,c)^\top\psi(i))$. However, successful conditioning\cite{liu2024crm,dai2021poso} is challenging in two aspects. First, we need to carefully pick or construct the user interest conditions. While explicit interests are usually those selected by users at sign-up or later on, implicit interests need to be mined from user engagement sequence $s_u$. 
Second, we need to associate the positive samples $i$ to the right condition $c$ for the model to learn effectively. This can be done at model training time for implicit user interest modeling and at data logging time for explicit user interest modeling, which will be detailed in the following sections. We concentrate on condition construction and condition association as two key elements that illuminate the commonalities among the methods.

\subsection{Implicit User Interest Modeling}
\label{sec:dcm}
%We start from a review of capsule networks in recommendations~\cite{sabour2017dynamic-capsule,li2019multi-mind,chai2022user-umi} and then introduce Differentiable Clustering Module (DCM), our best-performing solution for implicit user interest modeling, which can be seen as a modified Capsule Networks. 

\subsubsection{Capsule Networks for Recommendations}
Li et al. proposed MIND~\cite{li2019multi-mind} to use Capsule Networks in recommendation systems to extract user interest representations from user engagement history. Capsules are groups of neurons whose embedding  vectors represent spatial hierarchies of the encoded concepts~\cite{sabour2017dynamic-capsule}. Each item $e_i$ in the user sequence can be seen as lower capsules, which are used to derive the higher capsules $\{c_j\}_{j=1}^{K_{im}}$ representing multiple user interests via dynamic routing~\cite{sabour2017dynamic-capsule}. Unlike its application in computer vision, Capsule Networks for recommendations have only one routing layer and the bilinear mapping matrix $S$ is shared among the capsules~\cite{li2019multi-mind,lu2024mind360,cen2020controllable-comirec,chai2022user-umi}, resembling a clustering procedure.

%We briefly recap the routing process from a clustering point of view, where we see the higher capsules as cluster centroids. 
First, Capsule Networks instantiate potential cluster centroids $c_j$ via Gaussian random initialization~\cite{sabour2017dynamic-capsule,li2019multi-mind}. 
At each routing step, the routing weights from $e_i$ to $c_j$ are derived as
\begin{equation}\label{eq:logits}
    b_{ij} \leftarrow \frac{exp(c_j^\top Se_i)}{\sum_k exp(c_k^\top Se_i)},
\end{equation}
and then update $c_j$ as a weighted sum of all the item representations with closer items having higher contributions,
\begin{equation}
    c_j \leftarrow squash(\sum_i b_{ij}Se_i)
\end{equation}
with
\begin{equation}
    squash(v)=\frac{||v||^2_2}{1+||v||^2_2}\frac{v}{||v||_2}
\end{equation}
as a non-linear activation that enables the L2 norm of the vector to represent the existence of a cluster centroid. This procedure is repeated for several steps until cluster centroids converge.

\begin{figure}
    \centering
    \includegraphics[width=\linewidth]{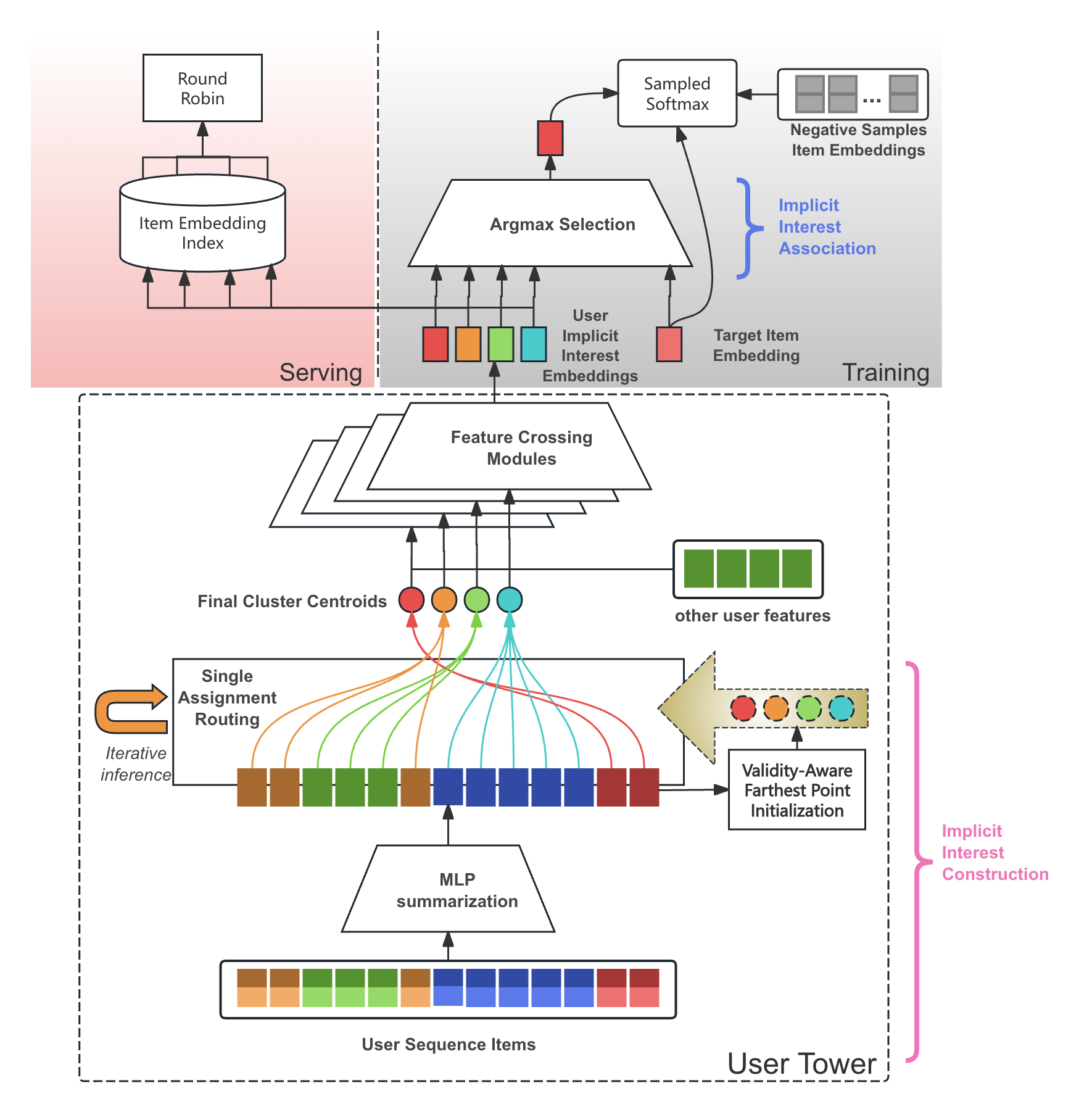}
    \caption{Implicit user interest modeling with the Differentiable Clustering Module. It differs from vanilla Capsule Network in the Validity-Aware Farthest Point Initialization and Single Assignment Routing to construct the implicit interest conditions. Condition association is performed via the argmax selection. We use these embeddings to retrieve candidates and do a round-robin merge at serving time.
    }
    \label{fig:dcm}
\end{figure}

\subsubsection{Implicit Condition Construction}
In this section, we describe the implicit interest condition construction method we adopt, namely \textbf{Differentiable Clustering Module (DCM)} as shown in Figure \ref{fig:dcm}. It differs from MIND~\cite{li2019multi-mind} in terms of the initialization and routing procedure, which we deem essential to clustering quality. 

At Pinterest, we use a set of features for user engaged items, including pretrained features~\cite{ying2018graph-pinsage} and categorical inputs. Different from MIND~\cite{li2019multi-mind} that pools and linearly projects the inputs with $S$, we learn the item representation $e_i$ from $N$ input features $f_{in}$ ($n$ denotes the $n$-th input item level feature) via an MLP summarization layer activated by $GELU$~\cite{hendrycks2016gaussian-gelu}:
\begin{equation}
    e_i=W_2^\top(GELU(W_1^\top\cdot concat(f_{i1},f_{i2},\ldots,f_{iN})))
\end{equation}
where $W_1$ and $W_2$ are weights for a 2-layer MLP. 

Initialization is known to be a crucial component for clustering algorithms to converge. We term our initialization heuristic as \textbf{Validity-Aware Farthest Point Initialization}, which ensures great coverage and diversity of the cluster centroids. Farthest Point Initialization (FPI)~\cite{arthur2006kmeans,wu2021centroid} assumes that given $m$ intialized cluster centroids $\{c_j\}_{j=1}^m$, we select the next cluster centroid iteratively as the item indexed by
\begin{equation}
\label{eq:maxmin}
	i^*={\arg\min}_i \max_j^m c_j^\top e_i.
\end{equation}
We first randomly select an item $e_i$ as the first cluster centroid and repeat the above procedure $K_{im}$ times until we have all the cluster centroids. Equation \ref{eq:maxmin} ensures that the initialized centroids are diversified as each iteration tries to minimize the maximum similarity between the next centroid and any given centroids.

However, it is important to incorporate validity filtering in the industry setup. In practice, certain input features significantly influence the output item embeddings. When these features are missing or deemed invalid, the resulting output embeddings can fall out of distribution. This misalignment critically affects our clustering algorithm, as the initialization process may incorrectly choose out-of-distribution items as cluster centroids. Without validity filtering, we observe significant regression on model performance and retrieval quality. With the validity filtering, Equation \ref{eq:maxmin} can be rewritten with a validity indicator  $\mathbf{I}_{valid}$ that masks out invalid items if they miss any feature inputs or indicate negative actions:
\begin{equation}
i^*={\arg\max}_i\min_j^m -\mathbf{I}_{valid}(e_i)c_j^\top e_i
\end{equation}

To further encourage the cluster centroids to diverge from each other after initialization, we incorporate the \textbf{Single-Assignment Routing} mechanism~\cite{lu2024mind360}. Recall the vanilla routing procedure in Equation \ref{eq:logits} assigns each item $e_i$ to every cluster centroid $c_j$ by a non-zero routing weights $b_{ij}$. When two cluster centroids are in close proximity, nearby items may contribute similarly to both centroids, potentially causing them to converge even further over time. 
Here we describe the single-assignment routing procedure from Lu~\cite{lu2024mind360}, which diverges the cluster centroids effectively by simply masking out non-maximum entries in the routing weights:

\begin{equation}
    b_{ij} \leftarrow \mathbf{I}(c_j^\top e_i=\max_k c_k^\top e_i) \frac{exp(c_j^\top e_i)}{\sum_k exp(c_k^\top e_i)}
\end{equation}

We provide extensive ablation study and visualization in Section \ref{sec:ablation} to show the importance of each component of DCM.

\subsubsection{Implicit Condition Association}
While the above sections describe condition construction for implicit user interests, we need to associate the target item to its corresponding interest. Given the $j$-th output user embedding $o_u^j=\phi(u,c_j)$ and the item embeddings $o_{y_i}=\psi(i)$, we achieve the condition association by our training objective, which involves one single user embedding that maximizes affinity to the target item in the sampled softmax loss ~\cite{bengio2003quick,yi2019sampling}.

\begin{equation}\label{eq:argmax}
    j^*={\arg\max}_j {o_u^j}^\top o_{y_i}
\end{equation}
\begin{equation}
    \mathcal{L}_{SSM}=-log\frac{exp({o_u^{j^*}}^\top o_{y_i}-logp_{y_i})}{exp({o_u^{j^*}}^\top o_{y_i}-logp_i)+\sum_{k\in \mathcal{B}}exp({o_u^{j^*}}^\top o_{y_k}-logp_{y_k})}
\end{equation}
where $y_i$ is the target item, $y_k$ represents in batch negative samples from $\mathcal{B}$, $p_{y_i}$ and $p_{y_k}$ represent streaming frequency estimation~\cite{yi2019sampling} for $y_i$ and $y_k$. 
Different from UMI\cite{chai2022user-umi}, we only use $o_u^{j^*}$ to compute the similarity for negative samples for computational efficiency, as the embedding layers and feature crossing modules are memory-intensive and occupy significant GPU resources.
%We choose $argmax$ for DCM according to MIND~\cite{li2019multi-mind}, which states that "model performs better with higher $p$". Notably, $argmax$ can be seen as a label-aware attention~\cite{li2019multi-mind} with $p=\infty$.

\subsubsection{Other Implicit User Interest Modeling Approaches}
\label{sec:others}
We also surveyed and explored a wide range of existing methods for implicit interest modeling at Pinterest, including self-attention~\cite{kang2018self-sasrec,cen2020controllable-comirec}, interest token based~\cite{xu2022mixture-mvke}, and PinnerFormer subsequence (PFS)\footnote{More details on PFS are included in the Appendix.} embeddings. We briefly introduce our implementation and share the empirical learnings  from online results.\\

\noindent \textit{Condition Construction.} As introduced in Section \ref{sec:related}, we adopt the multi-head self-attention module~\cite{cen2020controllable-comirec} to construct the implicit interest conditions for the self-attention method. For interest token based approach, we omit the multi-task setup and remove the virtual kernel gating in MVKE~\cite{xu2022mixture-mvke}, making the condition construction process equivalent to applying a single-head transformer decoder over the user sequence with learnable query interest tokens. 

For PFS, we generate multiple user embeddings by splitting user sequences into subsequences of different topics, which are generated by clustering 10 million random Pins into $K=32$ clusters using K-means on their Pinsage~\cite{ying2018graph-pinsage} embeddings. Then we assign sequence items to different topics based on their Pinsage similarity to the $K$ cluster centroids and group them by their assigned topics. This results in at most $K$ subsequences per user, which are then processed by the PinnerFormer~\cite{pancha2022pinnerformer} user sequence model to generate different user embeddings\footnote{More details in the Appendix.}. These embeddings can be used directly to retrieve candidates related to the topics, or as implicit conditions in a retrieval model. Empirically, using these subsequence embeddings as conditions performs better than using them for direct retrieval.\\

\noindent \textit{Condition Association.} Unlike the Dynamic Routing approach, if we use the $argmax$ operator to associate the positive item to conditions constructed by self-attention and interest token based approaches, the conditions will often collapse during training and the model will degenerate into a single-embedding retrieval model. We argue that when the interest condition representation involves random initialization, using a deterministic $argmax$ for embedding selection induces a “winner‐takes‐all” effect, where only one embedding receives meaningful gradients and the shared parameters collapse all embeddings into near-identical representations. The Dynamic Routing approach, however, constructs the condition representation by clustering the user sequence. When coupled with the modifications we have on DCM, it results in more diverse conditions by construction and prevents the model from collapsing. To prevent embedding collapsing for self-attention and interest token based models, we use a Straight-Through Gumbel Softmax~\cite{jang2016categorical-gumbel} to enable gradient flow to all implicit interest representations, which is the key to their convergence.

\begin{figure}[tbp]
    \centering
    \includegraphics[width=\linewidth]{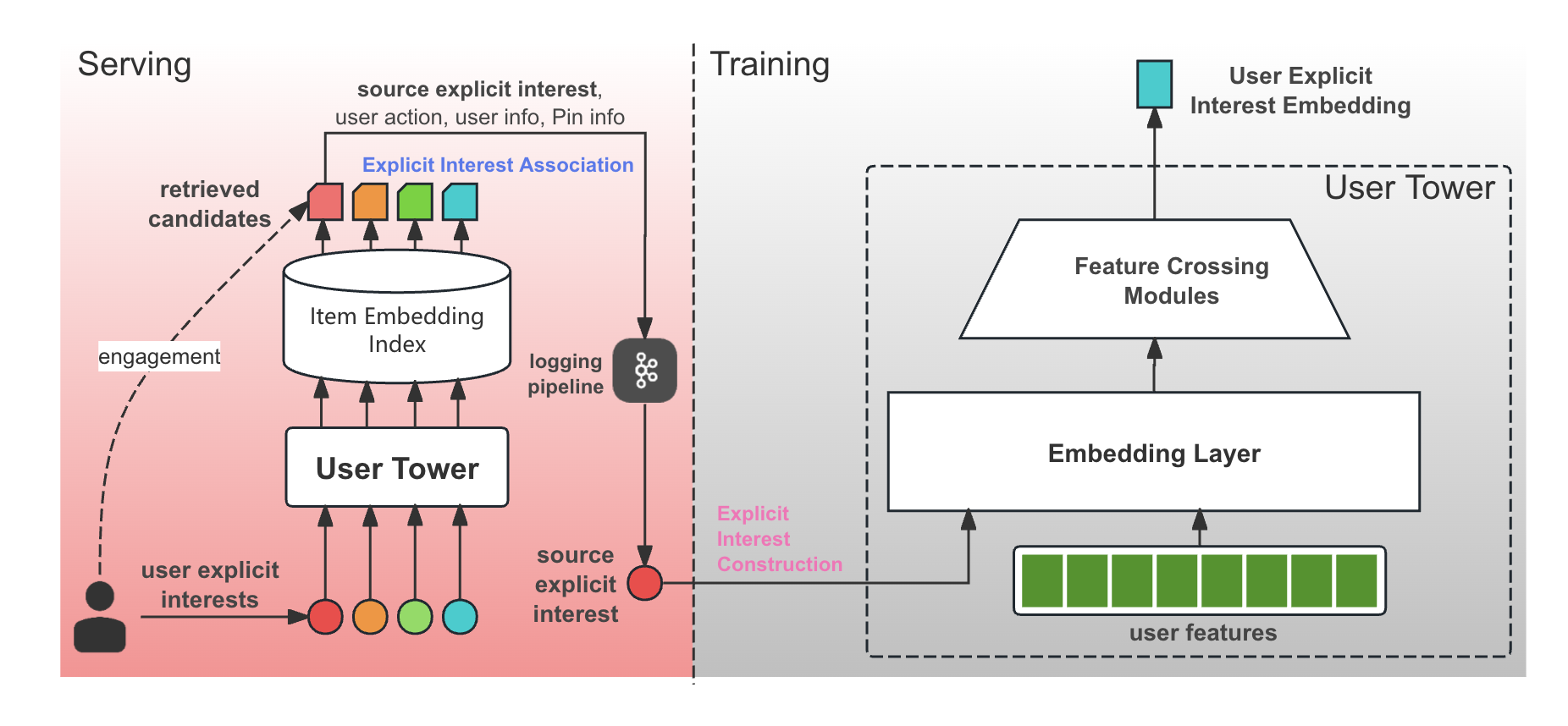}
    \caption{Explicit user interest modeling with Conditional Retrieval. The condition is constructed by embedding explicit signals (e.g., followed topics). Condition association is performed at serving time by tracking the source interests of engagements, which serve as the conditional inputs for training.
    }
    \label{fig:clr}
\end{figure}

\subsection{Explicit User Interest Modeling}
In addition to the implicit user interests extracted from user engagement sequence, we incorporate explicit user signals, such as followed topics, into our framework. Similar to many social media platforms, Pinterest users can follow topics (e.g., food, pets, and women fashion) during sign-up, with the flexibility to add or remove them at any time. Followed topics serve as clear indicators of user interests, which can be leveraged in retrieval.

In this section, we describe how we leverage Conditional Retrieval (CR)~\cite{lin2024bootstrapping} for explicit interest modeling. Similarly, CR can be broken down into condition construction and association steps. By leveraging conditional user engagement data, we can perform better at explicit interest modeling, as shown in the "Explicit Interest Association" path in Figure \ref{fig:clr}.

% These explicitly followed topics provide great business value for user understanding, as they may come from a longer-term history that cannot be directly fed into the model as raw user sequence, or they may represent interests that a user explicitly selects at sign-up but are gradually forgotten by the system. In addition, these explicit signals are valuable for inferring the needs from less active users, even more effective than implicit interest modeling according to our experiments. 

\subsubsection{Explicit Condition Construction} We modify the user tower by incorporating the condition information directly into the embedding layer. The condition embedding is then processed through feature crossing layers, allowing the model to capture higher-order feature interactions between the user and the condition.

\subsubsection{Explicit Condition Association} In Lin et al.~\cite{lin2024bootstrapping}, the goal is to retrieve items related to a given topic using only generic user-item engagement data. To bootstrap new use cases, they leverage existing item-to-topic signals and randomly sample a topic as the condition for the user tower. Essentially, the condition association is performed at training time. While effective for bootstrapping, this approach comes with a limitation: as one item can belong to multiple topics, the sampled topic may not overlap with users' followed topics or true intents, creating potential gaps between training and serving.

In the Pinterest home feed, we previously employed an inverted index retriever that fetches items based on users' followed topics. We collect user engagement data powered by this retriever and form training data (condition, user, item) tailored for explicit interest modeling. In other words, our condition association is performed at user action logging time, ensuring better alignment with users' followed topics. With the superior performance of CR, we subsequently replaced the inverted index retriever while keeping the same condition association process at logging time.

\subsubsection{Explicit Relevance Filter}

Ensuring condition relevance for explicit interest modeling is crucial for exposing users to their expected topics and improving the quality of the training data for explicit interest modeling. While CR already retrieves highly relevant items among embedding-based retrievers~\cite{lin2024bootstrapping}, we enhance this further by applying post-filtering based on item-to-topics signals as a guardrail for condition relevance, resulting in a considerable increase in engagements.

\subsection{Connections between Implicit and Explicit User Interest Modeling}
% To cover more interests at the retrieval stage, our multi-embedding retrieval framework leverages explicit interests to complement implicit interests from the user engagement history, since the user engagement history is usually capped to short-term engagements when fed into the retrieval model and explicit user signals may provide additional information on under-explored contents. 
% Hongtao's version
Implicit and explicit interest models complement each other at retrieval stage. Implicit interests are learned from user engagement sequence. Due to challenges in serving super long sequences within the retrieval model, implicit interest modeling usually captures shorter term interests. Explicitly followed topics complement the implicit ones in the following ways: 1) They may provide additional long-term and/or long-tail interests. 2) They are especially useful for less active users with limited user sequences. 3) They are more controllable and explainable, making it easy to inject heuristics to the retrieval pipeline. In practice, the average overlap of candidates retrieved from explicit and implicit interests conditioned user embeddings is \textbf{only 3.2\%}, indicating strong complementarity within our framework.

In terms of modeling techniques, they share the same philosophy but differ in implementation. The crux of the two models lies in two critical components: condition construction and accurate condition association. In terms of condition construction, the implicit interest modeling utilizes DCM to extract conditional signals through the clustering of user sequences. In contrast, explicit interest modeling involves the straightforward embedding of auxiliary input conditions. Regarding condition association, implicit interest modeling achieves this internally through the application of the argmax operator, while explicit interest modeling accomplishes this during the data logging phase. Both models effectively reframe the ranking score from 
$f(i|u)$ to $f(i|u, c)$.

\subsection{Deployment}
\label{sec:deploy}

For implicit interest modeling, we infer $K_{im}$ implicit user interests from the user engagement sequence. The corresponding user embeddings are used to perform ANN search from a per-computed item index. We assign different retrieval budgets based on the sum of their cluster routing weights $\sum_ib_{ij}$, interpreted as cluster importance. This prevents over-fetching candidates from torso and tail interests that deteriorate online performance. For explicit interest modeling, we randomly sample $K_{ex}$ followed topics and assign equal retrieval budgets for each conditional user embedding.

% At serving time, for both models, we combine each implicit or explicit interest condition with the rest of the user features and feed them into two separate models. We generate user embeddings and use them to do ANN search from a pre-computed item index in parallel. 
The retrieved candidates for each user embedding are then merged using a round-robin strategy with deduplication, ensuring a balanced mix of candidates from different interest conditions. This approach prevents over-representation of any single interest and delays ranking decisions to later stages for more refined personalization.

%For implicit interest modeling, we arrange the embeddings according to the sum of their cluster routing weights $\sum_ib_{ij}$ in descending order, and assign different candidate budgets based on the order. This prevents over-fetching candidates from torso and tail interests that deteriorates online performance. For explicit interest modeling, we apply post-filtering based on explicit interest conditions after ANN retrieval. This relevance guardrail provides auxiliary user impression increase from online experiments.

Since each user tower is only computed once per request, the increased model complexity is acceptable for online serving. Empirically, p90 latency for the embedding-based retrieval increases from 150ms to 205ms, remaining below responsiveness limits. In addition, to balance the serving cost and user engagements, we also reduce the number of candidates to retrieve per embedding and keep an on-par budget as the single embedding retrieval.

\section{Experiments}
In this section, we present our extensive experiments on the proposed multi-embedding retrieval framework, including comparisons and ablations of implicit and explicit interest modeling through both offline and online experiments. Additionally, we provide a visual case study to analyze the effectiveness of each component.

\subsection{Experiment Setup}
\subsubsection{Dataset and Metrics.} Following previous work~\cite{xia2023transact}, we compile the training dataset from a 15-day window of user engagement logs, using the first 14-day window for training and 15th day for evaluation. The dataset contains 6 billion engagement records from 160 million users.
For retrieval tasks, we formulate a composite prediction task that incorporates multiple positive actions, such as clicks, repins (saves), etc. Label sampling and weighting are applied to align with business objectives, but we omit the details for brevity to maintain the paper's focus.

For offline evaluation, we construct a retrieval corpus with 1 million of the most engaged items from the dataset. Each record in the evaluation dataset contains a single positive item to retrieve, and we measure performance using hit rate (\textbf{HR}) at various ranking thresholds. To compute the score of an item with multiple embeddings, we follow previous work~\cite{li2019multi-mind,cen2020controllable-comirec} to take the maximum score of each item among all user embeddings. In practice, since the positive item may not always be present in the constructed corpus, we compute HR by comparing the score of the positive item against the ranked items within the corpus. 

For online metrics, we report relative improvements on: (1) \textbf{HF Repins}, which is the repin (save) volume on home feed and serves as one of the most important metrics for this surface~\cite{xia2023transact}, and (2) \textbf{Adopted Pincepts (A-Pincepts)}\footnote{Pincepts is an internal annotation system to identify fine-grained interests for items. Adoption is defined as having 3 or more qualified actions in a week on a Pincept.}  as the online engagement diversity indicator. We also report them across user segments (core\footnote{We define users with >4 repins in 28 days as core users.} and non-core users).

\subsubsection{Implementation Details}
All models in the experiments are trained on the same dataset using sampled softmax loss with logQ correction~\cite{yi2019sampling} over the in-batch negatives. For both implicit and explicit interest modeling, the feature crossing layer in Figure \ref{fig:dcm} and Figure \ref{fig:clr} is a DHEN~\cite{zhang2022dhen} module that ensembles Transformers~\cite{vaswani2017attention},  Parallel Mask Net~\cite{wang2021masknet}, and MLPs. 
%For implicit interest modeling, the user sequence length is 200 with action filtering to focus on the most important engagements

In online comparison experiments, we fetch the same number of candidates and merge the candidates with round-robin before passing them to the ranking model, unless otherwise specified. We set $K_{im}$ = 7, $K_{ex}$ = 5, and the overall retrieval budgets for implicit and explicit models are $O(1k)$.

\subsubsection{Competitor Methods}
\label{sec:baseline}
Since our framework consists of two models, we conduct separate comparisons for implicit and explicit interest modeling against existing methods. For implicit interest modeling, we evaluate DCM against representative multi-interest retrieval methods from different modeling streams, as described in Section \ref{sec:others}, including MIND~\cite{li2019multi-mind}, self-attention~\cite{kang2018self-sasrec}, interest token-based~\cite{xu2022mixture-mvke}, and our in-house PinnerFormer Subsequence (PFS) based models. For explicit interest modeling, we report its performance comparison against inverted index-based retrieval, a widely adopted industrial solution for relevance-oriented retrieval, as well as CR with conditions directly extracted from item attributes~\cite{lin2024bootstrapping}.
\begin{table}
  \caption{Offline evaluation for implicit interest modeling.}
  \label{tab:dcm-offline}
  \begin{tabular}{ccc}
    \toprule
    Method&HR@100&HR@1000\\
    \midrule
    Self-Attention~\cite{cen2020controllable-comirec,kang2018self-sasrec} & 0.167& 0.470\\
    Interest Token~\cite{xu2022mixture-mvke} & 0.180& 0.474\\
    MIND~\cite{li2019multi-mind} & 0.175& 0.464\\
    DCM & \textbf{0.185}& \textbf{0.476}\\
  \bottomrule
\end{tabular}
\end{table}
\begin{table}
  \caption{Offline evaluation for explicit interest modeling.}
  \label{tab:clr-offline}
   \resizebox{\linewidth}{!}{
      \begin{tabular}{cccl}
        \toprule
        Method&filtered HR@100 
    & filtered HR@1000&HR@100 
    \\
        \midrule
        CR w/ item interest & 0.164 
    & 0.541&0.139
    \\
        CR w/ source interest & \textbf{0.191} & \textbf{0.565}&\textbf{0.145}\\
      \bottomrule
    \end{tabular}
}
\end{table}

\subsection{Offline Evaluation}
In this section, we report the offline comparison results for implicit and explicit interest modeling. 
Table \ref{tab:dcm-offline} presents the offline results for implicit interest modeling, where DCM outperforms the other methods in both HR@100 and HR@1000. The comparison between DCM and MIND~\cite{li2019multi-mind} also validates that the modifications upon Capsule Networks~\cite{sabour2017dynamic-capsule} in DCM improve the retrieval quality.

Offline results of differently trained Conditional Retrieval for explicit interest modeling are shown in Table \ref{tab:clr-offline}. We use different condition association strategies that either directly extract interest conditions from item attributes~\cite{lin2024bootstrapping} (termed as \textbf{CR w/ item interest}) or perform condition association at the time of action logging (termed as \textbf{CR w/ source interest}). Since explicit interest modeling applies term-based post-filtering at serving time, we replicate this step offline and report \textbf{filtered HR} to approximate the online treatment. Additionally, we include HR without filtering for completeness.  CR w/ source interest outperforms CR w/ item interest across all metrics, with a particularly strong advantage when filters are applied.

\begin{table}
  \caption{Online comparison for explicit interest modeling. Colored numbers (\cblue{blue} or \cred{red}) are statistically significant and \cgray{gray} numbers are non-statistically-significant.}
  \label{tab:online-clr}
  \resizebox{\linewidth}{!}{
      \begin{tabular}{cccc}
        \toprule
     control & inverted index & inverted index & item interest \\
     \cmidrule(lr{20pt}){1-4}
     enabled & CR w/ filter& CR w/o filter & source interest\\
        \Xhline{1pt}
        \rule{0pt}{1ex}
       HF Repins $\uparrow$ & \cblue{+0.56\%}& \cgray{+0.3\%}& \cblue{+0.98\%}\\
       NC-HF Repins $\uparrow$ & \cblue{+1.13\%}& \cgray{+0.46\%}& \cblue{+3.04\%}\\
       A-Pincepts $\uparrow$ & \cblue{+0.42\%}& \cblue{+0.37\%}& \cgray{+0.32\%}\\
       NC-A-Pincepts $\uparrow$ & \cblue{+0.44\%} & \cgray{+0.42\%} & \cblue{+1.03\%}\\
      \bottomrule
    \end{tabular}
}
\end{table}

\subsection{Online Experiments}
Given the well-known empirical challenge of offline-online metric discrepancy~\cite{krauth2020offline} especially for industrial retrieval models, we make design choices mainly based on online A/B testing for our framework. We obtain these metrics from experiments over a 2-week period.

We conduct online experiments on modeling and serving variants for explicit interest modeling in Table \ref{tab:online-clr}. We report metrics for all users and non-core users to showcase its strength on non-core users. We attach an "NC-" prefix in Table \ref{tab:online-clr} to indicate metrics over non-core users. 
% As the online experiments are not concurrent, we list out control and enabled groups in Table \ref{tab:online-clr}. 
Applying post-filtering for Conditional Retrieval (CR w/ filter) performs the best among other modeling or serving variants. When comparing against the same model without filtering (CR w/o filter), the notable gains demonstrate that enforcing interest relevance is helpful for this component, even when explicit context relevance is not a must for home feed. 
We also compare using the item interest from item attributes with the source interest from action logging as conditions. There is an increasing trend in all metrics if we use source interest, especially on HF Repins. It shows that appropriate condition association is crucial to improve engagements effectively. In addition, all metrics on non-core users are amplified, indicating that explicit interest modeling is adept at non-core users.

We then present the results for implicit interest modeling in Table \ref{tab:online-multi}. It is worth noting that these experiments are conducted with post-filtered CR launched to production. Thus, the metric gains in Table \ref{tab:online-multi} show a complementary effect of implicit interest-based retrieval. Consistent with offline results, DCM outperforms other methods in both HF Repins and A-Pincepts, while the self-attention-based method performs comparably. Notably, MIND~\cite{li2019multi-mind} relatively underperforms compared to other approaches without DCM’s modifications, further validating the effectiveness of centroid divergence in improving representation quality. We also report metrics for core users, as implicit interest modeling tends to benefit active users more due to its reliance on user sequence modeling. All methods show greater improvements on core users. Additionally, diversity metrics such as A-Pincepts exhibit strong correlation with HF Repins, reinforcing that improving diversity and user interest coverage at the retrieval stage benefits the overall system.

\begin{table}
  \caption{Online comparison for implicit interest modeling. }
  \label{tab:online-multi}
  \resizebox{0.9\linewidth}{!}{
      \begin{tabular}{ccccc}
        \toprule
        \multirow{2}{*}{Methods}&
        \multicolumn{2}{c}{HF Repins $\uparrow$}&
        \multicolumn{2}{c}{A-Pincepts $\uparrow$} \\
        \cline{2-5}
        \rule{0pt}{3ex}
     & all & core & all & core \\
        \Xhline{1pt}
        Self-Attention~\cite{cen2020controllable-comirec,kang2018self-sasrec} & \cblue{+0.83\%}& \cblue{+1.01\%}& \cblue{+0.44\%}& \cblue{+0.63\%}\\
        Interest Token~\cite{xu2022mixture-mvke} & \cblue{+0.68\%}& \cblue{+0.95\%}& \cgray{+0.21\%}& \cgray{+0.19\%}\\
        MIND~\cite{li2019multi-mind} & \cgray{+0.43\%}& \cblue{+0.56\%}& \cgray{+0.04\%}& \cgray{+0.15\%}\\
        PFS & \cblue{+0.47\%}& \cblue{+1.02\%}& \cblue{+0.32\%}& \cblue{+0.46\%}\\
        DCM & \cblue{\textbf{+0.86\%}}& \cblue{\textbf{+1.23\%}}& \cblue{\textbf{+0.46\%}}& \cblue{\textbf{+0.87\%}}\\
      \bottomrule
    \end{tabular}
}
\end{table}

\begin{table}
    \centering
    \caption{Online lift of the proposed framework.}
    \label{tab:framework}
    \begin{tabular}{c|ccc}
    \toprule
         Control &  Sitewide Repins $\uparrow$ &HF Repins $\uparrow$ &A-Pincepts $\uparrow$\\
         \midrule
         + Framework & \cblue{\textbf{+0.48\%}} & \cblue{\textbf{+1.09\%}} & \cblue{\textbf{+0.81\%}} \\
         \bottomrule
    \end{tabular}
    \label{tab:my_label}
\end{table}

\begin{table}
  \caption{Ablation Study on DCM Model Architecture.}
  \label{tab:ablation}
  \resizebox{0.9\linewidth}{!}{
      \begin{tabular}{ccccccl}
        \toprule
        VA-FPI & SAR & \#Cluster & HR@100 & HF Repins $\uparrow$ & A-Pincept $\uparrow$ \\
        \midrule
        $\times$ & $\times$ & 7 & 0.175& \cgray{+0.43\%} & \cgray{+0.04\%} \\
        $\times$ & $\checkmark$ & 7 & 0.176& \cgray{+0.37\%} & \cgray{+0.30\%} \\
        $\checkmark$ & $\times$ & 7 & 0.175& \cgray{-0.10\%} & \cgray{+0.17\%} \\
        $\checkmark$ & $\checkmark$ & 2 & 0.176& \cred{-0.53\%} & \cgray{+0.25\%} \\
        $\checkmark$ & $\checkmark$ & 4 & 0.180& \cgray{-0.15\%} & \cgray{+0.29\%} \\
        $\checkmark$ & $\checkmark$ & 7 &  \textbf{0.185}& \cblue{\textbf{+0.86\%}} & \cblue{\textbf{+0.46\%}} \\
      \bottomrule
      % \rule{0pt}{1pt}
    \end{tabular}
    }
\end{table}
\begin{figure}[h]
  \centering
  \includegraphics[width=0.8\linewidth]{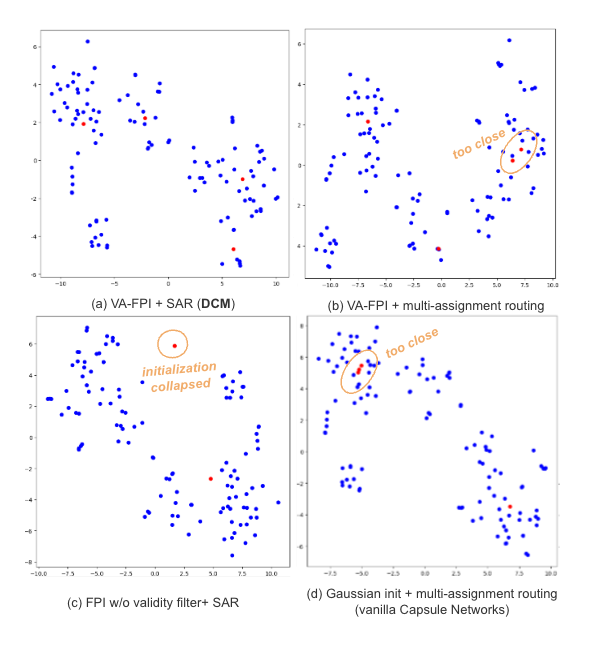}
  \caption{Comparison between DCM and Capsule Networks with 4 clusters. These embeddings are sampled from the same user and visualized with t-SNE, embeddings may locate differently across figures due to random projection.}
  \label{fig:capsule-viz}
\end{figure}

From the above discussion, we can see that implicit and explicit interest modeling show a complementary effect on different user segments. To measure the joint effect of the framework, we run a retrospective experiment and report the overall gains achieved by our framework in Table \ref{tab:framework}. Synergizing implicit and explicit interests increases feed diversity and user engagements by a large margin, including sitewide repins, which demonstrates cascading impact beyond the directly applied surface.

\begin{figure*}
    \centering
    \includegraphics[width=0.72\linewidth]{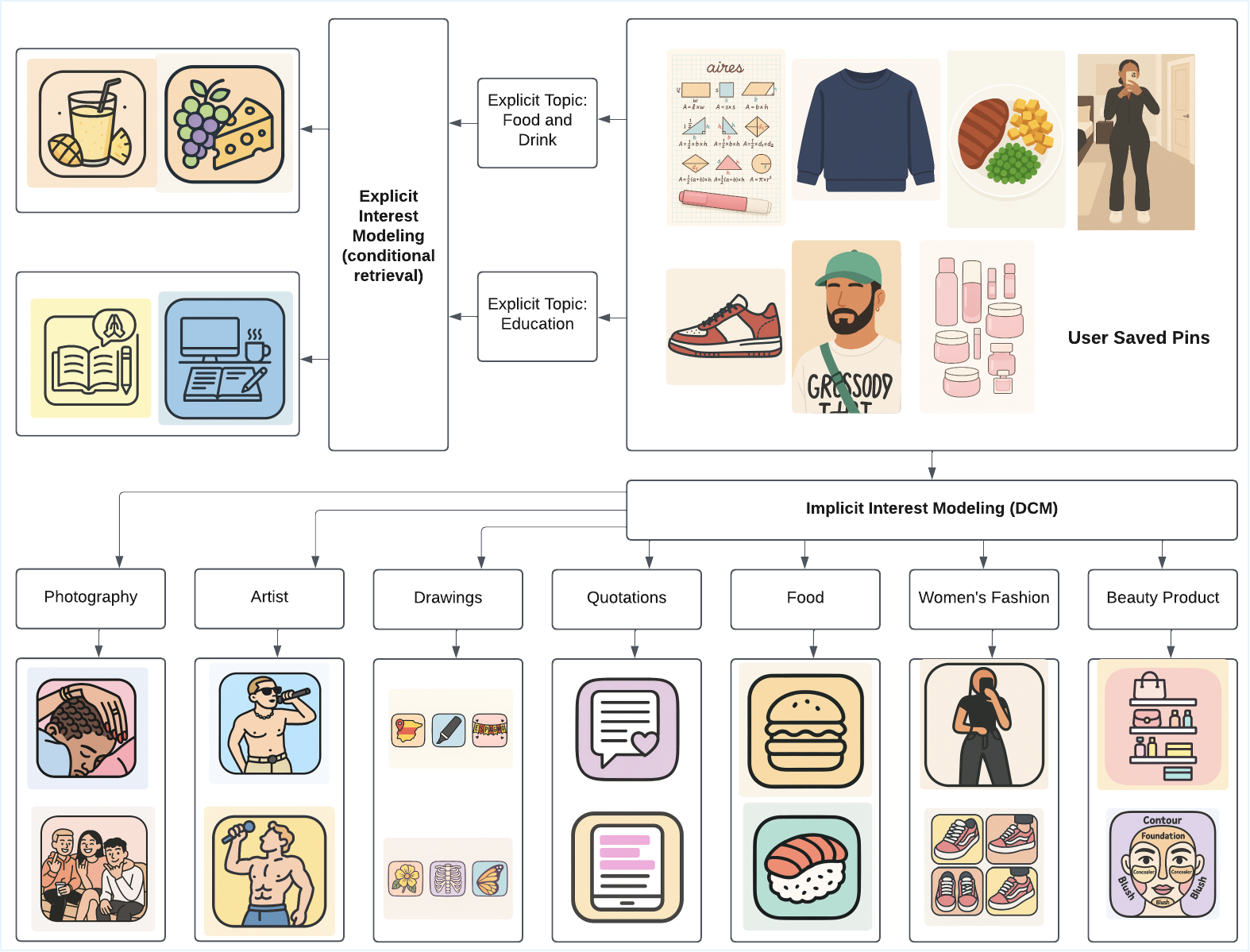}
    % \caption{Case study of a Pinterest user. The left section shows the user's saved Pins,  while the right section presents examples of retrieved candidates. Blue and yellow boxes highlight candidates from implicit and explicit interest modeling, respectively, with each box containing examples for one embedding. Candidates from implicit interest modeling are labeled with the most relevant interest term, while candidates from explicit interest modeling are labeled with the source interest. Best viewed in color with zoom-in. }
    \caption{Case study of a Pinterest user. For privacy reasons, we iconize the original images in the user history and recommendation feed to obscure identifiable user information. We show the user's saved Pins and recommendations from implicit interest modeling and explicit interest modeling separately, with each box containing 2 examples for one embedding. Candidates from implicit interest modeling are manually annotated with a common interest term within the retrieval set, while candidates from explicit interest modeling are labeled with the source interest.}
    \label{fig:case}
\end{figure*}

\subsection{Ablation Study}
\label{sec:ablation}
We run extensive ablation studies both offline and online to verify the design of DCM, including Validity Aware Farthest Point Initialization (VA-FPI), Single Assignment Routing (SAR)~\cite{lu2024mind360}, and number of clusters. As shown in Table \ref{tab:ablation}, the joint effect of VA-FPI and SAR on enhancing the cluster centroid divergence improves model capability and online metrics significantly. We also visualize the clustering procedure inside the neural network as shown in Figure \ref{fig:capsule-viz}. We take the PinSage~\cite{ying2018graph-pinsage} embeddings from the user sequence and extract 4 cluster centroids to inspect the clustering process. Without validity filtering, some cluster centoids will collapse to one point as shown in Figure \ref{fig:capsule-viz}(c). From Figure \ref{fig:capsule-viz}(a) to (d), by removing SAR~\cite{lu2024mind360} and VA-FPI, the cluster centroids are getting more concentrated, indicating suboptimal diversity to capture user interests holistically.

We also study the impact of different number of clusters. As shown in Table \ref{tab:ablation}, less number of clusters significantly deteriorates online performance, and even decreases metrics when there are only 2 clusters. This is possibly due to missing major interest coverage with less number of clusters. Similar to MIND~\cite{li2019multi-mind}, we try to reduce number of clusters adaptively for different users, but results in a -0.12\% change in HF Repins compared with serving a fixed number of clusters. In addition, taking the top ranked candidates among all embeddings~\cite{li2019multi-mind,cen2020controllable-comirec} is also experimented online, but leading to a marginal decrease in the number of repin users by 0.35\%. Given these results, we finalize our design choice as outlined in Section 3. 

\subsection{Case Study}
In Figure \ref{fig:case}, we present a case study based on a randomly selected user. For privacy reasons, we iconize the original images in the user history and recommendation feed to obscure identifiable user information. The user’s previously saved Pins are shown in the top right corner, and two example Pins retrieved from each embedding source are displayed separately for explicit interest modeling and implicit interest modeling. Note that this is the same user as in Figure \ref{fig:homefeed} where we have shown that single embedding retrieval fails to cover the user's interests holistically. Results from explicit interest modeling (CR) are labeled with the source interest provided to the model, while results from implicit interest modeling (DCM) are manually annotated by the Pins' common interest taxonomy. The case clearly illustrates that the interests reflected in the user’s saved Pins are well covered by our framework, and the retrieved candidates exhibit high relevance and personalization. Notably, while DCM overlooks the interest in "education", it is successfully recovered by CR, showcasing the supplementary effect of the components in our framework. In addition, DCM shows personalized interest granularity beyond human-defined interest taxonomy. For example, "beauty" related Pins from DCM range from make-up to cosmetic organization (coarse-grained), while "photography" Pins are more fine-grained with a common theme of "candid youthful moments and lifestyle".

\section{Conclusion}
In this paper, we introduce the multi-embedding retrieval framework at Pinterest, which synergizes Differentiable Clustering Module (DCM) for implicit interest modeling and Conditional Retrieval (CR) for explicit interest retrieval to capture diverse user interests and enhance user engagements. We frame both models as conditional representation learning, introducing distinct mechanisms for condition construction and association to better encode and leverage user interests as conditions. Our extensive offline and online experiments, including A/B testing at Pinterest, demonstrate significant improvements in user engagement and content diversity, validating the effectiveness of our framework in a real-world production environment.

\begin{acks}
We would like to say thank you to our colleagues - Jay Adams, Raymond Hsu, and Dylan Wang for the support and suggestions on this work. %And we appreciate the valuable insights shared by Lu\cite{lu2024mind360} in their blog, which have inspired our design choices.
\end{acks}

\bibliographystyle{ACM-Reference-Format}
\bibliography{sample-base}

\appendix
\section{Feature Crossing Module}
While we illustrate the condition construction and association for implicit and explicit interest modeling in details previously, here we elaborate the detailed architecture of the feature crossing modules in our retrieval model. Both implicit and explicit interest modeling share the same architecture in each tower. 

Our feature crossing module is based on DHEN\cite{zhang2022dhen}, an ensembling framework that can parallelize and stack arbitrary submodules for latent feature interactions. We use a DHEN with two hierarchies:
\begin{itemize}
    \item 1st hierarchy: a 2-layer Transformer encoder\cite{vaswani2017attention} (256 hidden dims, 4 heads), and a 2-layer MLP (1024 hidden dims) in parallel
    \item 2nd hierarchy: a 4-block parallel mask net\cite{wang2021masknet} (128 hidden dims, 0.5 projection ratio), and a 2-layer MLP (1024 hidden dims) in parallel

\end{itemize}
The output of the parallel submodules in each hierarchy are summed and feed into the next layer. The input features are splitted and projected into feature fields with equal dimensions to facilitate field-wise feature interaction in a transformer, while the MaskNet and MLPs take in the concatenated feature fields.

\section{PinnerFormer Subsequnce (PFS)}
Here we elaborate the implementation of our in-house Pinner-Former\cite{pancha2022pinnerformer}
 Subsequence (PFS) model as an alternative implicit interest modeling approach that we explored. The core idea is, rather than taking in a full user history sequence into the model, PFS takes in a subsequence that belongs to a similar concept and generate an embedding given each subsequence for retrieval. We first collect a set of $O(1M)$ items in an offline workflow and run K-means clustering based on PinSage\cite{ying2018graph-pinsage} to get 32 distinct concept clusters. For the items in each user history sequence, they are then mapped into these 32 clusters and splitted into subsequences accordingly. At training time, the model is trained using the same DenseAllAction\cite{pancha2022pinnerformer} loss except that the input subsequence corresponds to the target item's concept cluster. 

While these embeddings generated from PFS can be directly used to serve as a retrieval model, however, we found that empirically serving a two-tower model with these subsequence embeddings as input performs better. Due to the serving complexity for the deep Transformer architecture of PFS, we populate the PFS outputs as a user signal that can be taken in as a feature for the two-tower models.  The two-tower model is trained in the same way as described in Section\ref{sec:dcm}, except that the DCM outputs being replaced with these PFS embeddings. We hypothesize the reason DCM performs better than PFS is due to (1) the clustering of PFS is not end-to-end trained such that the embeddings for the items cannot adapt to the model's training objectives, (2) the signal is not consumed in a real-time manner, and (3) the clusters are pre-defined concepts, whereas the cluster granularity in DCM can be adaptive to user history.

\section{Non-core User Performance Breakdown}
Here we provide a detailed breakdown in Table\ref{tab:non-core-breakdown} of non-core user performance lift by the explicit interest modeling to justify its effectiveness across non-core user cohorts. Note that new users only take up less than 10\% of the total non-core user segments, therefore the results here mainly illustrate the positive trend of user experience as they do not surpass a 0.05 p-value threshold due to the small population. These results provide a granular view of the real-world impact on non-core users and demonstrate the effectiveness of explicit interest modeling on users of less activities.

\begin{table}[h!t]
    \centering
    \caption{Detailed breakdown of explicit interest modeling across non-core user cohorts.}
    \begin{tabular}{c|cccc}
    \toprule
         User cohort &  Casual & Marginal &Resurrected &New\\
         \midrule
         HF Repins & +1.12\% & +1.46\% & +1.00\% &+0.56\%$^*$\\
         A-Pincepts & +0.23\% & +0.05\% & +0.67\% &+0.65\%$^*$\\
         \bottomrule
    \end{tabular}
    \label{tab:non-core-breakdown}
\end{table}

\end{CJK}  % End CJK block
\end{document}